\documentclass[mathleft,fleqn,%
]{an}
\usepackage{graphicx}
\usepackage[varg]{txfonts}
\begin{document}

\Pagespan{1}{}
\Yearpublication{2014}%
\Yearsubmission{2014}%
\Month{0}%
\Volume{999}%
\Issue{0}%
\DOI{asna.201400000}%


\title{Chemical abundance gradients from open clusters in the Milky Way disk: 
results from the APOGEE survey
}
\author{Katia Cunha\inst{1}\thanks{Corresponding author:
        {kcunha@on.br}},
Peter M. Frinchaboy\inst{2},
Diogo Souto\inst{1},
Benjamin Thompson\inst{2},
Gail Zasowski\inst{3},
Carlos Allende Prieto\inst{4},
Ricardo Carrera\inst{4},
Cristina Chiappini\inst{5},
John Donor\inst{2},
Anibal Garc{\'i}a-Hern{\'a}ndez\inst{4},
Ana Elia Garc{\'i}a P{\'e}rez\inst{4},
Michael R. Hayden\inst{6}, 
Jon Holtzman\inst{7},
Kelly M. Jackson\inst{2},
Jennifer A. Johnson\inst{8},
Steven R. Majewski\inst{9},
Szabolcs M{\'e}sz{\'a}ros\inst{10},
Brianne Meyer\inst{2},
David L. Nidever\inst{11},
Julia O'Connell\inst{2},
Ricardo P. Schiavon\inst{12},
Mathias Schultheis\inst{6},
Matthew Shetrone\inst{13},
Audrey Simmons\inst{2},
Verne V. Smith\inst{14},
Olga Zamora\inst{4}
}

\titlerunning{Metallicity Gradients from APOGEE Open clusters}
\authorrunning{K. Cunha for the APOGEE collaboration}
\institute{
Observat{\'o}rio Nacional - MCTI, Brazil
\and
Texas Christian University, USA
\and
NSF AAPF, Johns Hopkins University, USA
\and
Instituto de Astrof{\'i}sica de Canarias and Universidad de La Laguna, Spain
\and
Leibniz-Institut f\"ur Astrophysik Potsdam, Germany
\and
Observatoire de la Cote d'Azur, France
\and
New Mexico State University, USA
\and
The Ohio State University, USA
\and
University of Virginia, USA
\and
ELTE Gothard Astrophysical Observatory, Hungary
\and
University of Arizona, USA
\and
Liverpool John Moores University, UK
\and
McDonald Observatory, University of Texas, USA
\and
National Optical Astronomy Observatory, USA}

\received{XXXX}
\accepted{XXXX}
\publonline{XXXX}


\abstract{%
Metallicity gradients provide strong constraints for understanding the chemical
evolution of the Galaxy. We report on radial abundance gradients of Fe, Ni,
Ca, Si, and Mg obtained from a sample of 304 red-giant members of 29
disk open clusters, mostly concentrated at galactocentric distances
between $\sim$8--15 kpc, but including two open clusters in the outer disk. 
The observations are from the APOGEE survey. The chemical abundances were derived automatically
by the ASPCAP pipeline and these are part of the SDSS III Data Release 12.
The gradients, obtained from least squares fits to the data, are relatively flat, 
with slopes ranging from -0.026 to -0.033 dex kpc$^{-1}$ for the 
$\alpha$-elements [O/H], [Ca/H], [Si/H] and [Mg/H]
and -0.035 dex kpc$^{-1}$ and -0.040 dex kpc$^{-1}$ for [Fe/H] and [Ni/H], respectively.
Our results are not at odds with the possibility that metallicity ([Fe/H]) gradients are steeper
in the inner disk (R$_{GC}$$\sim$ 7--12 kpc) and flatter towards the outer disk. 
The open cluster sample studied
spans a significant range in age. When breaking the sample into age bins, there
is some indication that the younger open cluster population in our sample 
(log age $<$ 8.7) has a flatter metallicity gradient when compared with the gradients 
obtained from older open clusters.
}
\maketitle

\section{Introduction}
The chemical evolution of a galaxy depends on
many global variables; some of the more important ones being
the star formation history, the infall and outflow of gas, and the initial
mass function. These variables may differ from one galaxy to another, change
as the galaxy evolves, and, combined with the existence of more star formation in the center 
of the galaxy, can result, for example, in the overall radial decrease of the metallicity 
across the galactic disk (e.g. Chiappini 2002). Such metallicity gradients are commonly seen 
in spiral galaxies and represent important observational constraints 
for models describing the formation and the chemical evolution of a galaxy
(e.g., Minchev et al. 2014; Kubryk et al. 2015; Stanghellini et al. 2015). 

In the Milky Way, abundance gradients can 
be measured in a variety of populations, 
such as the young population of OB stars, H II regions and Cepheids; Planetary Nebulae; 
cool unevolved stars and red-giants in the field and in open clusters. The
red giants which are members of open clusters, in particular, offer the advantage that their 
ages and distances can be better constrained than for field stars. 
In addition, open clusters span a large range of ages, from Myr to Gyr, and can 
be used to gain insight into the time evolution of gradients in the Galactic disk. In this 
paper, we report on the radial abundance gradients of
Fe and Ni, as well as the alpha-elements O, Mg, Si and Ca 
from open clusters observed by the Apache Point Observatory Galaxy Evolution Experiment,
the APOGEE survey.

\section{Apogee observations}

The APOGEE survey (Majewski et al. 2015) was one of the four surveys in the
Sloan Digital Sky Survey, SDSS-III. The APOGEE multi-object spectrograph 
collected data during bright time between 2011 - 2014 and 
observed over 150,000 red giants from all stellar populations in the Milky Way. 
The APOGEE spectra have a resolution R= $\lambda$/$\delta$ $\lambda$ $\sim$ 23,000 
and spectral coverage between $\sim$1.5 -- 1.7 micron.

APOGEE targeted a large sample of stars in disk open clusters.
These observations constitute the Open Cluster Chemical Analysis and Mapping (OCCAM) survey,
a homogeneous and uniform dataset within the APOGEE survey that can be used 
to study abundance gradients. 
The sample discussed here contains 304 red-giants, 
which are members of 29 open clusters, covering galactocentric 
distances roughly between 7--23 kpc, but mostly concentrated between $\sim$8-15 kpc 
and with ages ranging from $\sim$200 Myr to 10 Gyr (Dias et al. 2002). 
A future paper describing the studied sample 
will be presented in Frinchaboy et al. (2016). 

\section{Results and discussion}

\subsection{Abundances and comparisons with the literature}

The chemical abundances and metallicities of the individual
stars in our sample were derived automatically by the 
APOGEE Stellar Parameters and Chemical Abundance Pipeline - ASPCAP
(Garcia Perez et al. 2015); these are part of the most recent 
SDSS-III Data Release, DR12
\footnote{available at https://www.sdss.org/dr12/}. 

%
Figure 1 shows comparisons of our average abundances for Fe, Ni, O, Ca, Si and Mg 
with results from other optical studies in the literature for the following clusters:
NGC 6819 (Bragaglia et al. 2001); 
Berkeley 29 (Carraro et al. 2004); 
NGC 6791 (Carraro et al. 2006; Bragaglia et al. 2014);
NGC 2243 (Jacobson et al. 2011a); 
NGC 2420, NGC 2158, NGC 7789, NGC 188 and M67  (Jacobson et al. 2011b). 
The mean differences between our results and the literature ($<$DR12--Literature$>$) and corresponding
dispersions 
for each element 
are indicated in the panels. For all elements, except Mg, the mean offsets are smaller than 0.1 dex.

\begin{figure}
\includegraphics[width=\linewidth,height=138mm]{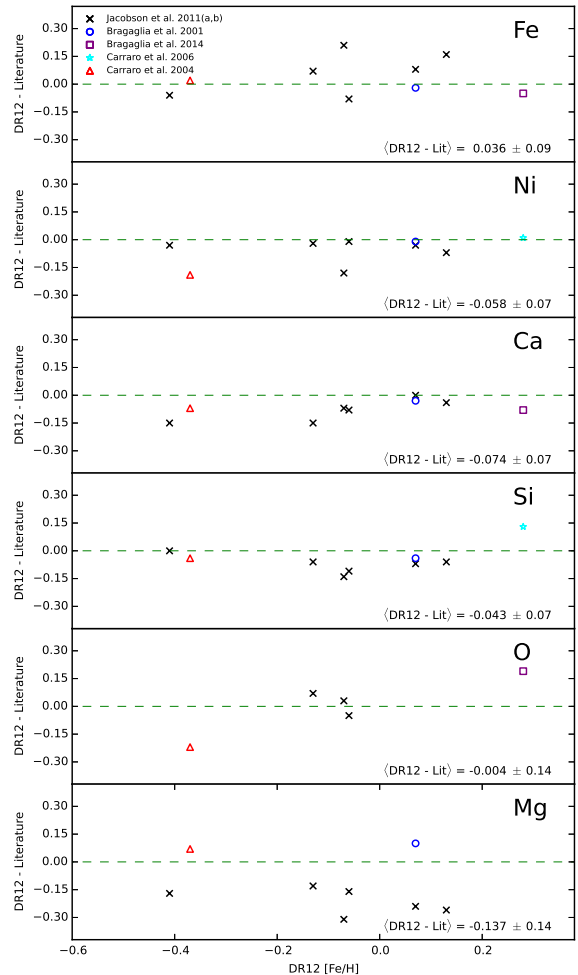}
\caption{A comparison of the ASPCAP/DR12 results with results from other optical
high-resolution studies in the literature. 
The selected clusters are calibration clusters for the APOGEE survey (M\'esz\'aros et al. 2103). 
The abundance differences shown in the y axis are the average abundances per
cluster and defined relative to the solar value as [el/Fe]$_{DR12}$ - [el/Fe]$_{Literature}$.
Typical uncertainties in the DR12 abundance averages are $\sim$$\pm$0.05 dex
(internal cluster abundance dispersions; Holtzman et al. 2015) and in the
differences 'DR12 - Literature' are estimated to be $\sim$$\pm$0.07 dex.
}
\label{label1}
\end{figure}

Overall, the average iron abundances from DR12 and the literature agree 
within $\sim$0.1 dex.
For two of the clusters, however, the derived metallicities are larger than 
those in Jacobson et al. (2011b) by $\sim$ 0.2 dex. 
The Ni, Ca and Si abundances for most of the clusters also agree with other 
studies within $\sim$ 0.1 dex, but with a tendency of  
being slightly lower in DR12 than in the literature. 
We note, however, that Carraro et al. (2006) find a higher Si abundance than ours for the very metal
rich cluster NGC 6791. 
For oxygen, there are fewer open clusters for comparison, but our results agree with those in 
Jacobson et al. (2011b), while there is +0.2 dex offset with the oxygen abundance in 
Bragaglia et al. (2014) for NGC 6791, and an opposite offset of -0.2 dex for Be 29,
when compared to Carraro et al. (2004).
The DR12 Mg abundances are systematically lower than those in Jacobson et al. (2011a,b),
but are slightly higher than in Bragaglia et al. (2001, 2014).
 
The overall conclusion of the comparison above is that the DR12 abundances do not seem to 
show significant systematic offsets when compared to these literature studies. 
Certainly, the abundance differences found are typical of what is generally seen 
in other comparisons of different abundance studies in the literature. In addition,
it is important to note that, since our goal is to investigate abundance gradients across the
Galactic disk, it is an advantage to have observational data and abundance analysis methods 
which are homogenous, minimizing the posibility of having spurious trends that could be due 
to systematics.
  
\subsection{Abundance gradients}

\begin{figure}
\includegraphics[width=\linewidth,height=168mm]{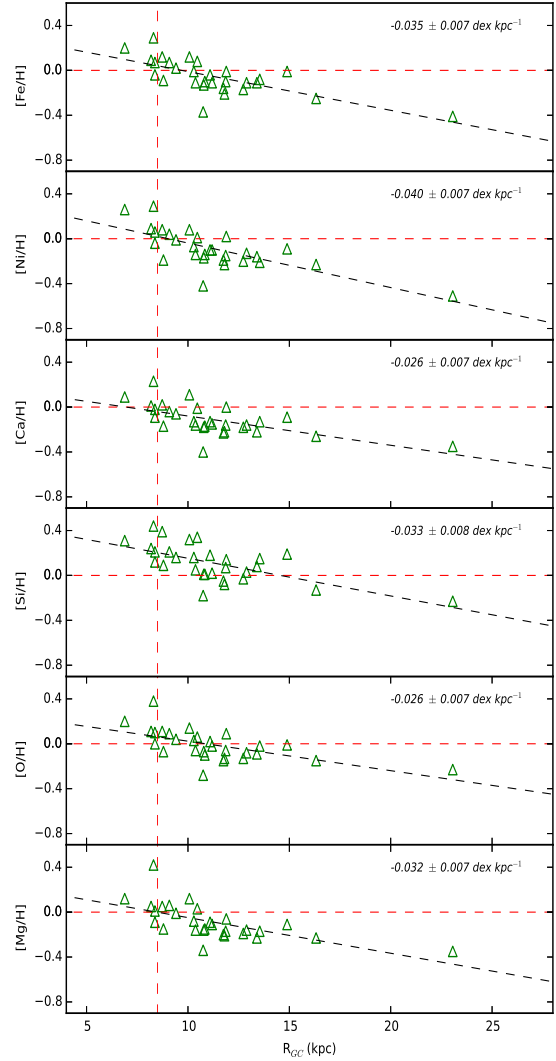}
\caption{Chemical Abundance gradients (best fit slopes) for a sample of open clusters observed 
by the APOGEE survey. The triangles represent average abundances for each cluster
computed from individual stellar abundances from cluster members in SDSS-III/ DR12. 
The galactocentric distances adopted for the open clusters are from the Dias et al. (2002) catalog.
}
\label{label2}
\end{figure}

An initial assessment of the metallicity (iron) and overall 
[$\alpha$/Fe] abundance gradients from the APOGEE OCCAM survey has been 
presented in Frinchaboy et al. (2013). This previous study was based on 
abundances in SDSS-III Data Release 10 (DR10; Ahn et al. 2013).
Other relevant studies of abundance gradients from open clusters in the literature
include, e.g., Donati et al. (2015); Yong et al. (2012); Pancino et al. (2010); Magrini et al. (2015).   

Figure 2 summarizes our results and abundance gradients obtained 
for Fe and Ni (products of Type Ia SN) and the $\alpha$-elements O, Ca, Si and Mg (products of Type II SN). 
In all panels of Figure 2 the green open triangles represent the average abundances 
obtained per open cluster. The typical internal abundance dispersions are $<$0.05 dex.
The horizontal and vertical dashed lines indicate, respectively, the solar abundance values 
and the galactocentric distance of the Sun (R$_{GC}$).
Overall, the abundance results for all elements, except Si, are roughly solar 
at close to solar galactocentric distances. In addition, the abundance scatter at roughly solar R$_{GC}$
is relatively small, but the open cluster NGC 6791 deviates from the average ([O/H] and [Mg/H] $\sim$+0.4 dex; see 
also Cunha et al. 2015).
The open cluster NGC 6791, however, is one of the most metal rich open clusters in the Galaxy and is
known to have some special characteristics such as being very old, very massive, lying
at $\sim$1 kpc from the galactic plane. 
As previously mentioned, for Si, our results are systematically higher 
than the solar Si abundance at roughly solar galactocentric distances. This would be an indication that our
Si abundances are overestimated, but the comparison with the open clusters in Figure 1 
indicates that our Si results are slightly lower than the literature.
We note, however, that DR12 results for [Si/Fe] are systematically higher than the results
in Bensby et al. (2014) by $\sim$0.1 dex.

Concerning the abundance gradients, best fit slopes and uncertainties are also presented in
Figure 2.  These were computed from least squares fits 
to the cluster average elemental abundances. 
When considering the entire sample the gradients obtained for [Fe/H] and [Ni/H] (mostly products 
of SN Type Ia) are -0.035 $\pm$ 0.007 dex kpc$^{-1}$ and -0.040 $\pm$ 0.007 dex kpc$^{-1}$, 
respectively, while the gradients for the $\alpha$-elements are slightly flatter with an 
average slope of -0.029 $\pm$ 0.004 dex kpc$^{-1}$. 
Previous studies have found evidence for a possible break in the metallicity gradients at R$_{GC}$$\sim$ 10--12 kpc
(e.g., Frinchaboy et al. 2013, Yong et al. 2012; Magrini et al. 2010), with a flatter gradient in the
outer disk when compared to the inner disk. 
When dividing our sample into clusters with R$_{GC}$$<$12 kpc and those with R$_{GC}$$>$12 kpc 
we also find that the metallicity gradient is flatter in the outer disk: -0.030 $\pm$0.009 dex kpc$^{-1}$,
while for the inner disk (R$_{GC}$$\sim$ 7--12 kpc) it is steeper: -0.068 $\pm$ 0.017 dex kpc$^{-1}$. 

Radial metallicity gradients from field stars in the APOGEE survey with distances from the Galactic
plane between 0.00$<$z$<$0.25 kpc, are found to be 
flat in the inner disk (R$_{GC}$$<$6 kpc) and quite steep (-0.087 $\pm$ 0.002 dex kpc$^{-1}$) between
R$_{GC}$$\sim$ 6--12 kpc, with a significant abundance scatter (Hayden et al. 2014). 
In the overlapping region from R$_{GC}$$\sim$7--12 kpc, the metallicity 
gradient obtained for the cluster sample (-0.068 $\pm$ 0.017) is somewhat flatter than for the field stars
based on DR10 results (these results should be revisited using DR12). 
We note, however, that the most distant cluster,
Be 29, has z=2 kpc and its metallicity is in line with the median metallicity found by Hayden et al. (2014) for
their sample of field stars with 1.00$<$z$<$2.00 kpc. (See also Cheng et al. 2012; Boeche et al. 2014)

The open clusters in our sample have considerable age spread 
($\sim$200 Myr -- 10 Gyr) and it is possible that
the derived metallicity gradient, 
which was obtained from open clusters of all ages,
carries also the signature of its evolution as the Galaxy evolves. In addition,
there is also the effect of radial migration.
In order to investigate the possible time evolution of metallicity gradients, 
we break our open cluster sample into three arbitrary age bins.  
The top and middle panels of Figure 3 show, respectively, the young (log age $<$ 8.7) 
and intermediate age (8.7 $<$ log age $<$ 9.0) clusters, and the bottom panel shows the oldest clusters,
with ages over $\sim$1 Gyr. The latter includes the only open cluster in our sample beyond 
R$_{GC}$$\sim$16 kpc (Be 29).
Best fit slopes are: -0.025 $\pm$ 0.017 and -0.037 $\pm$ 0.018 for the young and 
intermediate age populations, respectively. 
For the oldest clusters, we derive a steeper gradient of -0.049 $\pm$ 0.017
when considering the range R$_{GC}$$\sim$ 7--16 kpc, which is covered in all 3 age bins 
and flatter when including Be 29 (-0.036 $\pm$ 0.010). 
These preliminary results
indicate that the radial metallicity gradients are flatter for the younger open clusters.


\begin{figure}
\includegraphics[width=70mm,height=90mm]{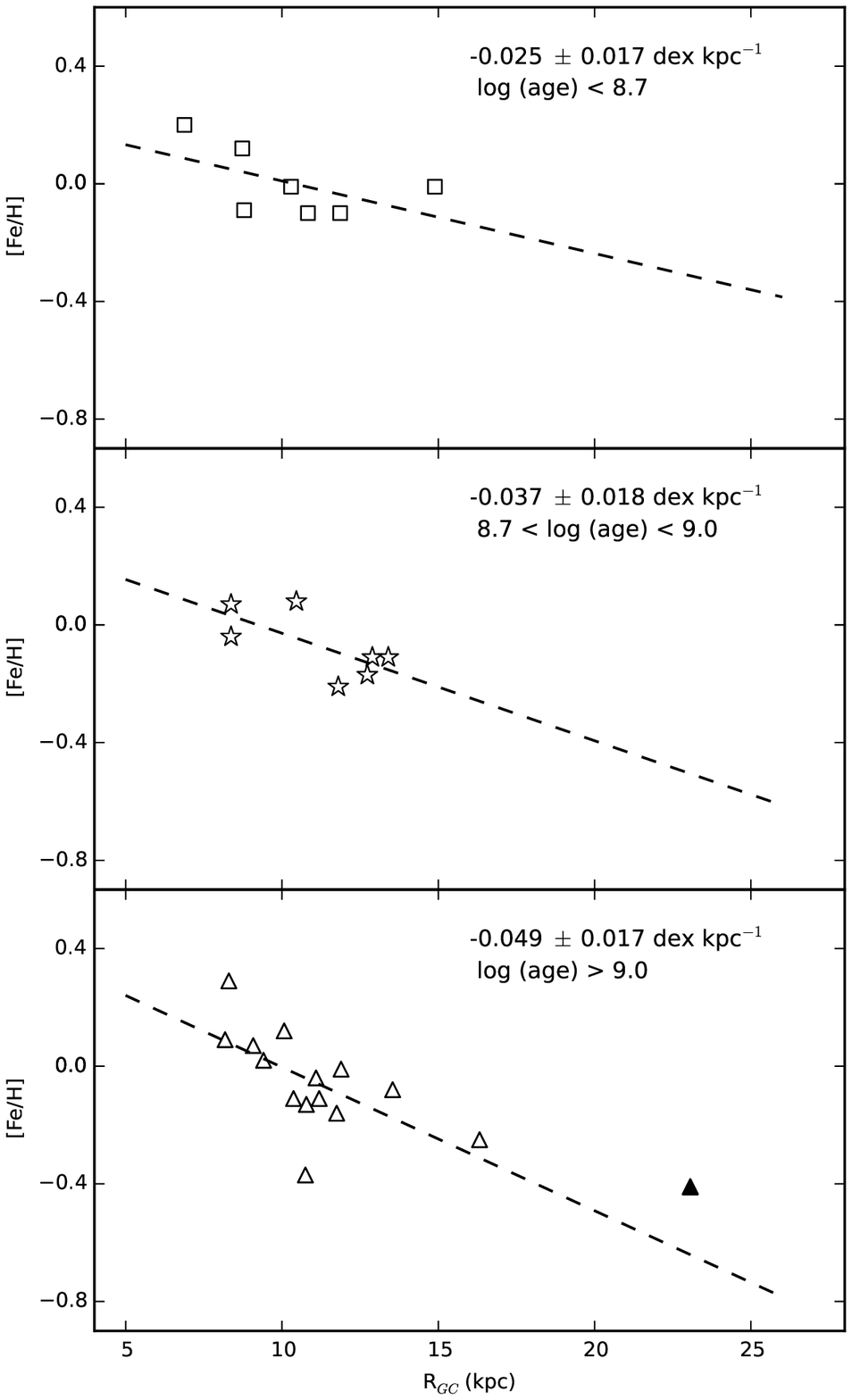}
\caption{The time evolution of gradients for the studied sample.
The open clusters in our sample are divided in 3 age bins: 
log age $<$ 8.7; 8.7 $<$ log age $<$ 9.0 and log age $>$ 9.0. 
The best fit slope for the old population did not include Be 29 (filled triangle).
The derived gradients are flatter for the youngest populations. 
}
\label{label1}
\end{figure}




\acknowledgements
PMF, BT, and JO are supported by NSF AST-1311835.
Funding for SDSS-III has been provided by the Alfred P. Sloan Foundation, the Participating Institutions, the National Science Foundation, and the U.S. Department of Energy Office of Science. The SDSS-III web site is http://www.sdss3.org/.
SDSS-III is managed by the Astrophysical Research Consortium for the Participating Institutions of the SDSS-III Collaboration including the Univ. of Arizona, the Brazilian Participation Group, Brookhaven National Laboratory, Carnegie Mellon University, University of Florida, the French Participation Group, the German Participation Group, Harvard University, the Instituto de Astrofisica de Canarias, the Michigan State/Notre Dame/JINA Participation Group, Johns Hopkins University, Lawrence Berkeley National Laboratory, Max Planck Institute for Astrophysics, Max Planck Institute for Extraterrestrial Physics, New Mexico State University, New York University, Ohio State University, Pennsylvania State University, University of Portsmouth, Princeton University, the Spanish Participation Group, University of Tokyo, University of Utah, Vanderbilt University, University of Virginia,University of Washington, and Yale University. 
  

%
%

\end{document}